\documentclass[journal]{IEEEtran}
\usepackage{comment}
\usepackage[english]{babel}
\usepackage{graphicx}
\usepackage{url}
\usepackage{cite}
\usepackage{booktabs}
\usepackage{multirow}
\usepackage{balance}
\usepackage{xcolor}
\usepackage{siunitx}
\usepackage{pgfplots}
\pgfplotsset{compat=newest}
\usetikzlibrary{plotmarks}
\usetikzlibrary{arrows.meta}
\usepgfplotslibrary{patchplots}
\usepackage{grffile}
\usepackage{amsmath}

\title{Remote IoT devices: sleepy strategies and
signal processing to the rescue for a long battery life}

\author{Gilles Callebaut, Geoffrey Ottoy, Guus Leenders, Bart Thoen, Lieven De Strycker, Liesbet Van der Perre}
\date{}

\begin{document}
\maketitle

\begin{abstract}
Long range wireless connectivity opens the door for new IoT applications. Low energy consumption is essential to enable long autonomy of devices powered by batteries or even relying on harvested energy. This paper introduces technologies for long-distance interaction with energy-constraint embedded devices. It proposes sleepy strategies for power management for the platform and the transmission. Adequate signal processing on the remote modules is demonstrated to play a crucial role to sustain the autonomy of these systems. The presented open-source development platform invites the signal processing community to a smooth validation of new algorithms and applications.
\end{abstract}

New wireless technologies serve IoT use cases in need of wide area coverage. Dedicated networks operating in unlicensed bands below 1~GHz, such as LoRaWAN~\cite{lora} and Sigfox~\cite{sigfox}, are being rolled out. Also new cellular communication modes and terminal categories are defined tailored for Machine Type Communication and IoT applications~\cite{3GPP}. This paper presents the landscape of different wireless technologies. The main concepts of the transmission schemes are introduced.

It is evident from electromagnetic wave propagation physics that long range and low power make contradictory requirements. We present a versatile IoT module with Low Power Wide Area Network (LPWAN) connectivity. It supports several sensors and performs real-time power monitoring. Our measurements, illustrated in Figure~\ref{fig:energy_profile}, show the transmitter (Tx) for a basic Lora communication to be up to 10 times more power hungry than the receiver (Rx). IoT traffic is mostly uplink dominated. Hence, the nodes predominantly need to provide the power demanding Tx. We propose slim strategies: put system components in sleep as much as possible and think before you talk (transmit). These sound and established design principles should be applied to the extreme in this new context.
\begin{enumerate}
\item In the \textbf{design} of the IoT platform evidently low power operation is key, yet sufficient processing means should  be supplied to enable local processing of captured data. Fine-grained power-off capabilities are vital on the platform. Select best-in-class components considering both dynamic and idle power.
\item Sleepy \textbf{operation} toggles individual subsystems and relies on an interrupt-based operation to wake-up when necessary. Adequate signal processing reduces, accumulates, and eventually interprets the data locally. This will result in fewer transmissions under better conditions.
\end{enumerate}

\begin{figure}[tb]
  \centering
    \includegraphics[width=0.45\textwidth]{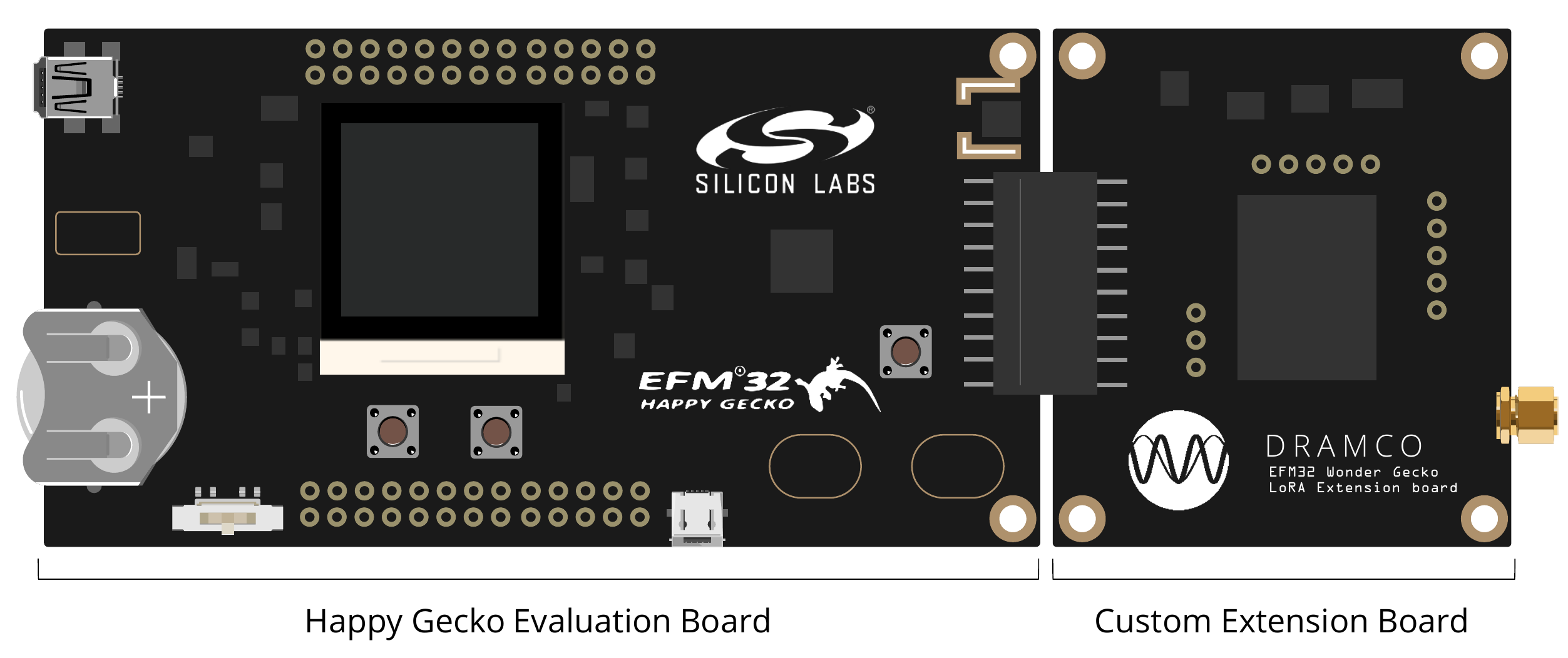}
    \caption{A Silabs STK3400 ``Happy Gecko'' together with a custom-design extension board creates a versatile IoT module that supports different sensors and performs real-time power monitoring. The software as well as the hardware is open-source: \protect\url{github.com/DRAMCO/LoRaWAN_EFM32}. }\label{fig:module} 
\end{figure}

Our design and experiments show that connecting remote IoT devices via an LPWAN is straightforward. Short time-to-validation, experimentation, and eventual deployment are in sight for many applications. Moreover, sleepy strategies and effective signal processing on the nodes in energy-constraint devices are elaborated.

The paper will include the following contributions that have not been published elsewhere and are relevant to the broad signal processing community:
\begin{itemize}
\item \textit{A total energy consumption breakdown of a typical sensor module connected via an LPWAN, including the communication, sensing, and processing portions.}
For the communication aspects, this trend is considerably more pronounced than what has been reported for shorter range and/or wider bandwidth technologies~\cite{bougard2007,NB_IoT}. Our assessment, based on real-life measurements, demonstrates the need and opportunity for adequate signal processing on the module to significantly extend battery life.
\item \textit{A versatile sensor module with real-life power monitoring capability, presented in conjunction with schematic and code in open source.}
The signal processing community can leverage on this convenient and accessible platform for easy validation of new algorithms and processing strategies.
\item  \textit{An illustrative case demonstrating an order of magnitude longer battery life for the LPWAN-connected IoT module through operating a slim strategy and clever power-optimized design.}
\end{itemize}

\section{Outline of Proposed Paper}

The following sections outline the planned structure for the proposed
survey paper. We summarize the relevant expertise of the
author-team in a further section, and link to open source materials provided complimentary to this paper. References include specific scientific papers as well as relevant introductory publications.

\subsection{Low Power Wide Area Connections: The Technological Landscape}
This section introduces the state of the art wireless communication technologies targeting the IoT. Complementary to other overviews~\cite{AdelantadoVTMM16}, this survey will assess the landscape of current IoT technologies considering energy consumption in conjunction with communication characteristics, based on different use case requirements.
The most prominent requirements for the IoT are long range and low power connections.
Consequently, this paper focuses on LPWANs which are tailored to provide wide range communication to power-constrained devices. LoRa~\cite{lora} and Sigfox~\cite{sigfox} are both LPWAN technologies which gained popularity in recent years and are now key players in their respective domain. These technologies mainly operate in the sub-GHz spectrum due to the favorable propagation characteristics needed for coverage extension. Furthermore, new cellular communication modes and terminal categories are defined adapted for Machine Type Communication and IoT applications~\cite{3GPP}. This section lays the foundation for the remainder of this paper.


\begin{figure}[tb]
  \centering
    \includegraphics[width=0.5\textwidth]{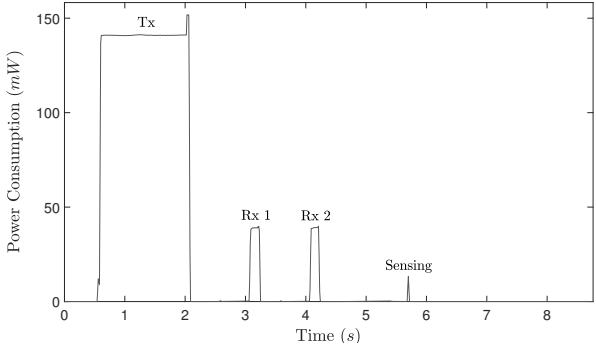}
    \caption{Energy profile (Tx, Rx and sensing) of our versatile LoRa board.}\label{fig:energy_profile}
\end{figure}

\subsection{Energy Profiling of an IoT Node reveals the transmitter big spender}

\begin{figure}[!ht]
  \centering
    \includegraphics[width=0.5\textwidth]{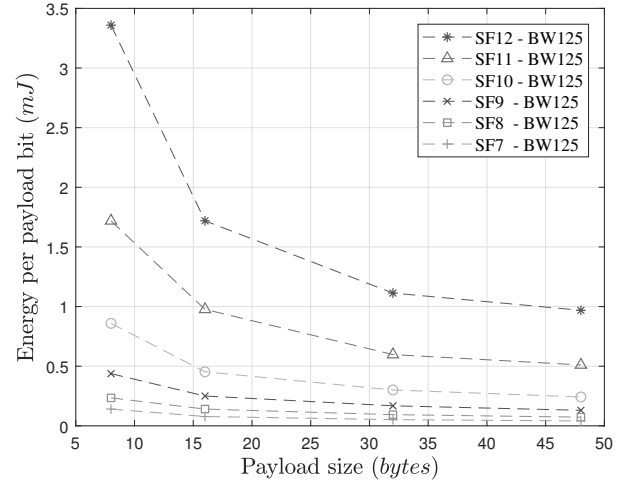}
    \caption{Energy consumption per bit reduces when more data is being transmitted at higher data rates in LoRaWAN.}\label{fig:energy_per_bit}
\end{figure}

The power dominance of wireless communication, and more specifically transmitting data, is illustrated in Figure~\ref{fig:energy_profile}.
The total energy consumption on the node will determine its battery life. 
As a consequence, we provide an overall assessment and breakdown of the energy consumption of a typical IoT node in various real scenarios.
This assessment includes measurements of the energy (Tx) per bit when transmitting messages at different data rates, as depicted in Figure~\ref{fig:energy_per_bit} 
The results clearly indicate the increased energy when using a lower data rate (i.e. more time on air). In addition, it also illustrates that accumulating data prior to transmitting it can increase the power efficiency of your device. These measurements allow us to determine the battery life of IoT nodes for a wide range of applications. 
Furthermore, these results provide a baseline for evaluating our sleepy strategies.

\subsection{Versatile IoT module facilitates low power development}
In this section the versatile IoT module with LPWAN capabilities, depicted in Figure~\ref{fig:module}, will be further elaborated upon. This versatile module supports a variety of sensors and performs real-time power monitoring. The module hosts an EFM32 Happy Gecko developer board. This board combines the powerful, yet low power, ARM Cortex M0+ with power sensing techniques. This makes the module an ideal development board for use in a very limited power budget system.

An IoT platform needs a way to communicate: a custom LoRa based extension board was built. This extension board features a LoRa transceiver, i.e. a Semtech SX1272 chip. This radio chip is combined with a variety of sensors: temperature, humidity, light, tactile, and energy consumption. This combination makes this module a truly versatile module for LPWAN IoT networks.

\begin{table*}[tb]
\centering
\caption{Energy consumption in \SI{}{\milli\joule} of Rx windows for different transmit DRs}
\begin{tabular}{@{}lcclcclccc@{}}
\toprule
\multicolumn{1}{c}{\multirow{2}{*}{\begin{tabular}[c]{@{}c@{}}Tx \\ DR\end{tabular}}} & \multicolumn{2}{c}{RX 1} &  & \multicolumn{2}{c}{RX 2} &  & \multicolumn{3}{c}{Total}                                                                                                   \\ \cmidrule(lr){2-3} \cmidrule(lr){5-6} \cmidrule(l){8-10} 
\multicolumn{1}{c}{}                                                                  & ACK       & NO ACK       &  & ACK       & NO ACK       &  & \begin{tabular}[c]{@{}c@{}}ACK\\ worst-case\end{tabular} & \begin{tabular}[c]{@{}c@{}}ACK\\ best-case\end{tabular} & NO ACK \\ \midrule
DR 0                                                                                  & -         & 6.4          &  & 5.6       & 1.3          &  & 12                                                       & -                                                       & 7.7    \\
DR 1                                                                                  & -         & 3.3          &  & 5.6       & 1.3          &  & 8.9                                                      & -                                                       & 4.6    \\
DR 2                                                                                  & -         & 1.6          &  & 5.6       & 1.3          &  & 7.2                                                      & -                                                       & 2.9    \\
DR 3                                                                                  & -         & 1.3          &  & 5.6       & 1.3          &  & 6.9                                                      & -                                                       & 2.6    \\
DR 4                                                                                  & 2.9       & 0.7          &  & 5.6       & 1.3          &  & 6.3                                                      & 2.9                                                     & 2.0    \\
DR 5                                                                                  & 1.7       & 0.5          &  & 5.6       & 1.3          &  & 6.1                                                      & 1.7                                                     & 1.8    \\ \bottomrule
\end{tabular}
\end{table*}

\subsection{Sleep While You Can: The one-and-only way to use your hardware right.}

It is clear that increasing the time a sensor spends in sleep mode, will increase the battery life. When focusing on the sleep mode of an IoT device: one should consider several design parameters.
First of all, one should select components (both sensors and microcontroller) with a low power drain in sleep mode. For example, the ARM Cortex M0+ offers several energy modes (EM) with a current ranging from \SI{150}{\micro\ampere\per\mega\hertz} in EM0 to \SI{20}{\nano\ampere} in EM4~\cite{emodes}. As a sensor example, we consider the APDS-9200 luminosity sensor, which has a typical standby current of \SI{1}{\micro\ampere}~\cite{adps-9200}. This is definitely low, however, when this sensor is not being used for some time, one should provide the circuitry to cut its power. A good example of this approach can be found in Thoen~\textit{et al.}~\cite{ThoenArray}.

A second important design goal should be to increase sleep time by working in an interrupt-based manner. Several applications only require action when something is erroneous. For instance, when the temperature in a forested area rises above a certain threshold, a fire alert needs to be triggered. In all other cases, we do not really care what the temperature is. To avoid a poll-based approach, where the device's controller reads out the temperature sensor to determine if it is above the threshold, one could use either digital sensors with programmable thresholds or use an analog sensor and one of the controller's built-in comparator modules. In the former approach, the digital sensor has an extra output pin that serves as an interrupt to the controller. The controller itself remains in sleep until it is interrupted. In the latter approach, only the comparator module remains active (with a certain threshold set). When the threshold is reached, the comparator module wakes up the controller core.

Making a distinction between relevant and irrelevant measurements is not only a way to save energy when it comes to sensing, it also saves energy when communicating.

\subsection{Think Before You Talk}

In a typical sensor node with wide area connectivity, the transmitted energy can be up to ten times higher than the energy needed to receive.  Hence, a persistent `think before you talk'  tactic is essential to minimize the overall energy consumption of the nodes. One strategy is accumulating nontime-critical sensor data. As a result, the overhead related to header information decreases when increasing the payload data. This effect is shown in Figure~\ref{fig:energy_per_bit}.
Furthermore, reducing the total number of transmission contributes to the minimization of the energy consumption in several other ways. First of all, it lowers the overhead of starting and initializing a transmission. It also reduces the number of retries (when necessary). Finally, it also reduces the number of downlink receive windows.

As discussed in the previous section, introducing processing to distinguish relevant and non-relevant information is imperative in power-constrained devices. Therefore, a `compute before you talk` is necessary to review data prior to transmitting it. Moreover, signal processing algorithms ought to be used to compress fast-chanching signals, i.e. to represent information with as few bits as possible.

Intelligent trade-off strategies, between transmit power and data rates, are exploited to further decrease the energy consumption of IoT nodes.
LoRaWAN defines such a mechanism to lower the transmission energy without increasing the packet error rate (PER).
This feature is called Adaptive Data Rate (ADR) where the data rates, airtime, and energy consumption in the network is optimized. Based on the history of the signal-to-noise ratio (SNR) and the number of gateways that received each transmission the transmit power or data rate can be increased; allowing to decrease the energy consumption of the node. This computation power needed for this feature is outsourced to the network.

\subsection{Conclusions and Future Directions}

The main messages and contributions of the paper will be summarized.
Specific conclusions that will be supported in the paper include:

\begin{itemize}
\item Connecting IoT devices through Low Power Wide Area Networks is easy, facilitated by open source tools and platforms. Hence, validation of new R\&D can be performed swiftly. The door is open to develop a wide variety of creative applications.
\item Representative sensors and a microcontroller can operate on less then \SI{1}{\micro\ampere}, when in sleep. When actively sensing, however, the current consumption can go up to even \SI{10}{\milli\ampere}. This shows the need for a keen sensing (and sleeping) strategy. Even more so, the transmit energy is at least an order of magnitude higher than the sensing energy. Hence, a persistent 'Think before you talk' tactic, i.e., deciding \textit{when and what to send}, is essential to minimize the overall energy consumption of the nodes.
\item Processing platforms for IoT nodes should enable fine-grained power-off modes. This can be achieved by a well-considered component selection and a thoughtful design of both hardware and software of the device, thus allowing meticulous power management.

\end{itemize}


Promising paths are identified for future exploration of further energy reductions. A specifically relevant approach is to equip the access point with a smart antenna array, potentially operating a Massive MIMO system~\cite{Larsson2014}. It can reduce the transmit power in the devices significantly while providing simultaneous access to many nodes.


Additional energy savings can be achieved by using custom-made wakeup chips or hardware. This approach has shown to be very effective for example in an acoustic wakeup with a passive microphone~\cite{AAD}.

As it is currently ``raining IoT technologies'', we can not expect a single communication technology to suffice for every use case. It would be interesting to develop a platform where multiple IoT technologies were combined to create a fusion of communication technologies. That way, the platform would be able to communicate with different networks for different payloads, energy levels, urgency levels, etc. Also, this platform could further reduce the energy consumption of IoT nodes by intelligently switching between technologies.

\section{Authors' Expertise and Resources related to the paper}

The authors constitute a complementary team with expertise both on energy efficient wireless systems~\cite{EARTHcommag}\cite{GreenRadioSPSmag} and on building low power embedded systems. Real-life experiments have been conducted with the modules presented in this paper in the DRAMCO lab of the KU Leuven, focusing on the Design and Research of Embedded Connected Systems. Connectivity has been established both to external gateways and to one installed on the campus.

The authors have recently run tutorials related to this paper, entitled `Low power wireless technologies for connecting embedded sensors in the IoT: A journey from fundamentals to hands-on.` at IEEE sensors 2017 and PIMRC 2017. They perform original research in this domain involving Ph.D. students and post-doctoral researchers. The lab performs research at the forefront of 5G communication~\cite{larsson2017massive}  and low power sensor-platforms~\cite{ThoenArray}. Furthermore, they proliferate this know-how on topical technologies in their related courses on mobile communications and embedded systems.

The author-team fosters a culture of sharing knowledge and creating accessible technological solutions. All documentation, code, and schematics for the presented versatile sensor module with LPWAN connectivity are made available in open source through a dedicated website (\url{dramco.be/tutorials/low-power-iot/}). This knowledge base welcomes the R\&D community a smooth validation of technological innovation and a large range of IoT applications.

\bibliographystyle{IEEEtran}
\bibliography{citations}

\begin{IEEEbiographynophoto}{Gilles Callebaut} is pursuing a Ph.D. in Massive MIMO for low power Machine Type Communication (MTC). He initiated the tutorial `Low power wireless technologies for connecting embedded sensors in the IoT: A journey from fundamentals to hands-on' and is currently a co-instructor to further enhance the tutorial.
Gilles graduated summa cum laude in 2016 and received the M.Sc. degree in engineering technology at KU Leuven campus Gent, Belgium. In addition, he received the laureate award. He is currently a member of DRAMCO, a research group which is focused on wireless and mobile communication systems. His main interests are Machine Type Communication (MTC), Internet of Things (IoT), low power embedded systems and everything mobile.
\end{IEEEbiographynophoto}
\balance
\begin{IEEEbiographynophoto}{Geoffrey Ottoy} is currently working as a post-doctoral researcher on acoustic localization using swarm networks and location-based services in Internet of Things applications. His main research topics are indoor localization and embedded wireless design. Since 2006 he has written several articles and collaborated in research projects about these subjects. He combines his research with teaching activities on topics such as digital and embedded design.

He received the Master degree in Engineering Technology (Electrical Engineering) from the Catholic University College Ghent, Belgium, now part of KU Leuven, in 2006. In 2013 he received his Ph.D. in Electrical Engineering from KU Leuven, Belgium.
\end{IEEEbiographynophoto}
\vspace{-0.8cm}
\begin{IEEEbiographynophoto}{Guus Leenders} is a member of the DRAMCO research group of the Electrical Engineering Department of the KU Leuven. He is involved with numerous of projects in the context of IoT, cooperating closely with industry including SMEs. His research is complemented by teaching activities. His main interests are Internet of Things and embedded systems. Guus received his master’s degree summa cum laude in engineering technology at KU Leuven (campus Ghent, Belgium) in 2015. There, he received the laureate award as well.
\end{IEEEbiographynophoto}
\vspace{-0.8cm}
\begin{IEEEbiographynophoto}{Bart Thoen} is currently working on his Ph.D. on ultra-low-power acoustic localization using wireless acoustic sensor networks (WASN). Since October 2013, he started as a researcher at the KU Leuven, ESAT department as a member of the DRAMCO (wireless and mobile communications) group, working on a project about wireless power transfer. He received the Master degree in Industrial Sciences: Electronic Engineering from the University College KAHO Sint-Lieven, Belgium, in 2013.
\end{IEEEbiographynophoto}
\vspace{-0.8cm}
\begin{IEEEbiographynophoto}{Lieven De Strycker} is a full professor at the Faculty of Engineering Technology, Department of Electrical Engineering, KU Leuven. He was a coordinator of European Erasmus Intensive teaching programs (Life Long Learning program). Prof. L De Strycker was an invited speaker on indoor localization at several universities (Darmstadt, Porto, Iasi, Plovdiv, Suceava, La Rochelle).

In 2001 he joined the Engineering Technology department of the Catholic University College Ghent (KAHO Sint-Lieven), where he founded the DRAMCO (wireless and mobile communications) research group in cooperation with ESAT-TELEMIC, KU Leuven. He is still the coordinator of the DRAMCO research group which has been involved in over 20 national and international research projects. He did an IAESTE traineeship, Siemens Madrid, Spain. He did research in European FP5 and national R\&D projects at the INTEC\_design Lab of Prof. Jan Vandewege and received the Ph.D. degree in Electrotechnical Engineering from Gent University in 2001 (summa cum laude). His Master degree in Electrotechnical Engineering he received at Ghent University, in 1996 (summa cum laude).
\end{IEEEbiographynophoto}
\vspace{-0.8cm}
\begin{IEEEbiographynophoto}{Liesbet Van der Perre} is Professor at the department of
Electrical Engineering at the KU Leuven in Leuven, Belgium and a guest
Professor at the Electrical and Information Technology Department at
Lund University, Sweden. Dr. Van der Perre was with the
nano-electronics research institute imec in Belgium from 1997 till
2015, where she took up responsibilities as senior researcher, system
architect, project leader and program director. She was appointed
honorary doctor at Lund University, Sweden, in 2015. She was a
part-time professor at the University of Antwerp, Belgium, from 1998
till 2002.  She received her Masters and a Ph.D. degree from the KU Leuven, Belgium,in 1992 and 97 respectively.

Her main research interests are in wireless communication, with a
focus on the physical layer and energy efficiency in transmission,
implementation, and operation.  Prof. L. Van der Perre was one of the initiators for the  European project EARTH focusing on Energy Aware Radio and Networking Technologies, and the scientific leader of FP7-MAMMOET, Europe's prime project on Massive
MIMO technology. She is a member of the Board of
Directors of the company Zenitel since 2015. Liesbet Van der Perre is
an author and co-author of over 300 scientific publications, including papers published in IEEE communication and signal processing magazine on energy efficient networks and green radios.
\end{IEEEbiographynophoto}


\vspace{\fill}

\end{document}